\documentstyle[11pt,aaspp]{article}
\catcode`\@=11 % This allows us to modify PLAIN macros.
\def\lsim{\mathrel{\mathpalette\@versim<}}
\def\gsim{\mathrel{\mathpalette\@versim>}}
\def\@versim#1#2{\vcenter{\offinterlineskip
        \ialign{$\m@th#1\hfil##\hfil$\crcr#2\crcr\sim\crcr } }}
\catcode`\@=12 % at signs are no longer letters
\eqsecnum
\begin{document}
\date{}
\title{Heating and Cooling of Hot Accretion Flows by Non Local Radiation}
\author{Ann A. Esin}
\affil{Harvard-Smithsonian Center for Astrophysics, Cambridge, MA 02138}

\begin{abstract}
We consider non-local effects which arise when radiation emitted at one 
radius of an accretion disk either heats or cools gas at other radii through 
Compton scattering.  We discuss three situations:

1. Radiation from the inner regions of an advection-dominated flow
Compton cooling gas at intermediate radii and Compton heating gas at
large radii.

2. Soft radiation from an outer thin accretion disk Compton cooling a
hot one- or two-temperature flow on the inside.

3. Soft radiation from an inner thin accretion disk Compton cooling
hot gas in a surrounding one-temperature flow.

We describe how previous results are modified by these non-local
interactions.  We find that Compton heating or cooling of the gas by the 
radiation emitted in the inner regions of a hot flow is not important.   
Likewise, Compton cooling by the soft photons from an outer thin disk is 
negligible when the transition from a cold to a hot flow occurs at a
radius greater than some minimum $R_{\rm tr,min}$.  However, if the hot flow
terminates at $R < R_{\rm tr,min}$, non-local cooling is so strong that the 
hot gas is cooled to a thin disk configuration in a runaway process. 
In the case of 
a thin disk surrounded by a hot one-temperature flow, we find that Compton 
cooling by soft radiation dominates over local cooling in the hot gas
for $\dot{M} \gsim 10^{-3} \alpha \dot{M}_{Edd}$, and $R \lsim 10^4 R_{Schw}$.
As a result, the maximum accretion rate for which an advection-dominated
one-temperature solution exists, decreases by a factor of $\sim 10$,
compared to the value computed under an assumption of local energy balance.
\end{abstract}


\section{Introduction}

Hot optically thin accretion flow solutions are often used to model
X-ray binaries and active galactic nuclei.  In addition to the
original two-temperature model of Shapiro, Lightman \& Eardley (1976)
and the corona model of Haardt \& Maraschi (1991), two new classes of
advection-dominated models have been proposed recently: a two-temperature 
solution (Narayan \& Yi 1995, Abramowicz et al. 1995), and a
one-temperature solution (Esin et al. 1996).  

Advection-dominated models differ from the usual accretion solutions 
in that only a fraction of the viscously dissipated energy is 
radiated locally in the disk.  The remainder of the generated 
energy is stored in the gas as entropy, resulting in a very hot, optically 
thin, quasispherical flow.  If the accreting object is a black hole, as it is 
believed to be the case for several X-ray binaries and all active galactic 
nuclei,
the stored energy is carried by the gas inside the horizon, so that the
luminosity of the system can be orders of magnitude smaller than is 
predicted by standard accretion theory.  The gas in advection-dominated 
solutions cools primarily through bremsstrahlung, synchrotron and inverse
Compton scattering processes.  Since the accreting electrons are heated to 
temperatures of up to $10^{10}\,{\rm K}$, the spectra of such systems span 
$\sim 10$ orders of magnitude in energy, from synchrotron radio emission
to $\sim 400 keV$ X-rays produced via Compton cooling of hot electrons.
Because two-temperature advection-dominated solutions have low radiative 
efficiency and hard power-law spectra, they have been 
successful in reproducing the observed properties of various 
low-luminosity X-ray sources including
Sagittarius A$^*$ (Narayan, Yi, \& Mahadevan 1995), the black hole 
nova A0620-00 in quiescence (Narayan, McClintock \& Yi 1996), and the central 
source in NGC 4258 (Lasota et al. 1996).  They also appear to be relevant 
to luminous systems (Narayan 1996).  It is not yet clear if one-temperature
solutions are relevant to any observed system.

Though detailed numerical spectra of advection-dominated flows include 
global radiative transfer effects (Narayan, McClintock \& Yi 1996; Narayan 
1996; Narayan, Barret \& McClintock 1997), analytical calculations of the 
physical properties of these accretion solutions, which we briefly review in 
\S2, generally ignore the effects of non-local radiative transfer (Narayan \& 
Yi 1995; Esin et al. 1996).  This is a dangerous approximation.  The optical 
depth for electron scattering in such flows is generally much 
lower than unity, and therefore, radiation emitted at one radius can 
potentially change the energy balance of the gas at quite different radii. 
It is the purpose of this paper to investigate the importance of this effect.  
We analyze three different but related issues.

First, in the advection-dominated models the bulk of the luminosity is emitted 
from the innermost region of the disk ($R < 100 R_{Schw}$), where most of the 
gravitational energy is dissipated and the cooling efficiency of the gas is 
higher.  The radiative flux from this region can have two effects on the rest 
of the flow, as we discuss in \S3.1.  The hot radiation can Compton heat 
(\S3.1.1) the outer regions ($R > 1000 R_{Schw}$), where the gas is cooler 
(Shvartsman 1971, Ostriker at al. 1976, Grindlay 1978).  On the 
other hand, cold but numerous synchrotron photons can contribute to cooling 
of the gas (\S3.1.2) closer in ($100 R_{Schw} < R < 1000 R_{Schw}$).   

Next, the hot solutions based on the one-temperature advection model (Esin 
et al. 1996) do not extend down to the last stable orbit  for
$\dot{M} > 0.003 \alpha^2 \dot{M}_{Edd}$.  At accretion rates higher than this 
limit the only possible configuration is a standard thin disk 
(Shakura \& Sunyaev 1973; Frank, King \& Raine 1992), surrounded by a hot 
flow.  Since most of the energy is dissipated in the thin disk, the number 
density of the
cold photons emitted in the inner regions can be much greater than the 
number density of photons emitted locally by the hot gas, and we expect that 
the hot flow will suffer significant extra cooling via Compton scattering of
the thin disk photons.  This issue is discussed in \S3.2. 

Finally, both one- and two-temperature advection-dominated solutions can be 
surrounded by a thin accretion disk, with the boundary between them at
some radius $R_{tr}$, where the value of $R_{tr}$ depends on the 
parameters of the model.  Here again the cold photons emitted by the thin disk
can Compton cool electrons in the nearby hot flow.  We analyze this situation 
in \S3.3.

Just as the radiation from the thin disk affects the energy balance of the 
hot flow, the cold gas itself is affected by irradiation from the 
advection-dominated zone.  In \S3.4 we investigate whether this process
affects our conclusions from \S3.2 and \S3.3.
We summarize and discuss the results in \S4.  

\section{Advection-Dominated Accretion Flows} 

In this section we review the basic equations, derived by Narayan \& Yi (1995),
that describe local properties of advection-dominated accretion flows in terms
of five basic parameters: mass of the accreting object, $M$, mass accretion 
rate, $\dot{M}$, radial distance from the accreting object, $R$, standard 
viscosity coefficient, $\alpha$ (Shakura \& Sunyaev 1973), and the ratio of 
gas pressure to total pressure, $\beta$.   

\subsection{Self-Similar Solution}

The total pressure in a vertically averaged, axisymmetric accretion flow
is given by
\begin{equation}
\label{prho}
p = \rho c_s^2,
\end{equation}
where $\rho$ is the gas density and $c_s$ is the local isothermal sound speed.
We assume that the gas is in equipartition with a tangled magnetic field, so
the total pressure is the sum of the gas and magnetic pressures, 
$p = p_g + p_m$.  In the ionized gas, the expression for the gas 
pressure is 
\begin{equation}
\label{pg}
p_g = \beta \rho c_s^2 = \frac{\rho k T_i}{\mu_i m_u} + 
\frac{\rho k T_e}{\mu_e m_u},
\end{equation}    
where $T_i$, $T_e$, and $\mu_i$, $\mu_e$ are temperatures and molecular 
weights of ions and electrons respectively.  The magnetic pressure due to
the isotropic tangled magnetic field is given by 
\begin{equation}
\label{pm} 
p_m = (1-\beta) \rho c_s^2 = \frac{B^2}{24 \pi}.
\end{equation}
Note that this equation differs from the one adopted by Narayan \& Yi 
(1995) by a factor of $1/3$.

Narayan \& Yi (1995) have derived a self-similar solution for 
advection-dominated accretion flows that obey mass, energy, and momentum
conservation.  Combining this solution with the equation for the gas pressure, 
and scaling mass in units of the solar mass, $M = m M_{\odot}$, accretion rate
in Eddington units, $\dot{M} = \dot{m} \dot{M}_{Edd} = \dot{m}\ 1.39
\times 10^{18} m\ {\rm g\,s^{-1}}$, and radii in
Schwarzschild units, $R = r R_{Schw} = r\ 2.95 \times 10^5 m\ \mbox{cm}$, we 
obtain 
\begin{equation}
\label{ne}
n_e = 1.33 \times 10^{19} \frac{\dot{m}}
{m\alpha (c_3 r)^{3/2}}\ {\rm cm^{-3}},
\end{equation}
\begin{equation}
\label{q+v}
q^+_v = 1.84\times 10^{21}\ \frac{\epsilon^{\prime} \dot{m} \sqrt{c_3}}
{m^2 r^4}\ {\rm erg\,cm^{-3}\,s^{-1}},
\end{equation}
\begin{equation}	
\label{T}
T_i + 1.08 T_e = 6.66\times 10^{12} \frac{\beta c_3}{r}\ {\rm K},
\end{equation}
where $n_e$ is the electron density, $q^+_v$ is the rate of viscous energy 
dissipation per unit volume, and we have taken $\mu_i = 1.23$ and 
$\mu_e = 1.14$, as required for a gas composition of 75\% H and 25\% He.  
The parameters $c_3$ and $\epsilon^{\prime}$ are defined as
\begin{equation}
\label{c3}
c_3 = \frac{2 (5+2 \epsilon^{\prime})}{9 \alpha^2} \left\{\left[1+
\frac{18 \alpha^2}{(5+2 \epsilon^{\prime})^2}\right]^{1/2} - 1\right\}
\end{equation}
and
\begin{equation}
\label{ep}
\epsilon^{\prime}=\frac{1}{f} \left(\frac{5/3 - \gamma}{\gamma-1}\right) = 
\frac{1-\beta}{f};
\end{equation}
where $f$ is the fraction of viscously dissipated energy that is
stored in the gas, and the expression for the adiabatic exponent $\gamma$ is 
derived in Appendix A.  

\subsection{Heating and Cooling Processes and Local Energy Balance}

\subsubsection{Heating of Electrons}

Since ions are considerably more massive than electrons, viscous heating 
affects primarily the ions (Shapiro, Lightman, \& Eardley 1976; Phinney 1981; 
Rees et al. 1982).  However, cooling of the hot plasma occurs mainly 
through electrons, so the energy has to be transferred in some way from 
ions to electrons.  In their analysis, Narayan \& Yi (1995) (following 
Shapiro, Lightman, \& Eardley [1976]) assumed that the energy transfer 
can occur only through Coulomb interactions at a volume transfer rate, 
$q_{ie}$, the expression for which was derived by Stepney \& Guilbert 1983.  
Since $q^+_v \propto \dot{m} \propto n_i$ (Eq. [\ref{q+v}]) and 
$q_{ie} \propto n_i n_e$ (Stepney \& Guilbert 1983), at low densities, 
Coulomb interactions are less 
efficient than viscous energy dissipation, giving rise to two-temperature 
accretion flows with ions much hotter than electrons.  

In addition to Coulomb scatterings, 
some non-thermal coupling mechanisms have been 
proposed recently (e.g. Begelman \& Chiueh 1988).  If such mechanisms are 
more efficient than Coulomb coupling, they can restore the thermal equilibrium 
between the ions and electrons, creating a one-temperature accretion flow
(Esin et al. 1996).  In that case, the question of energy transfer from ions
to electrons is settled by simply setting $T_i = T_e$ at all times. 

\subsubsection{Local Cooling Processes}

The fraction $(1-f)$ of the energy generated by viscous heating must be 
radiated by the gas. 
Cooling of the hot plasma occurs via three processes: bremsstrahlung,  
synchrotron radiation, and inverse Compton scattering of low energy 
bremsstrahlung and synchrotron photons:
\begin{equation}
\label{q-}
q^- = q^-_{br} + q^-_{sync} + q^-_{C}.
\end{equation}

\paragraph{Bremsstrahlung and Line Emission}
The free-free emission in a plasma is produced in electron-electron and 
electron-ion interactions.  The  corresponding cooling rates 
were derived by Stepney \& Guilbert (1983) and Svensson (1982).
The total bremsstrahlung cooling rate per unit volume is the sum of the two 
contributions, $q^-_{br} = q^-_{e e}+ q^-_{e i}$.

At large radii, the gas is cold enough to allow line cooling, which becomes
more efficient than free-free emission below $10^8\ {\rm K}$.  At these 
temperatures electron-electron bremsstrahlung emission is negligible, and
we use a cooling function that includes electron-ion bremsstrahlung and line 
cooling (updated version; original version was described by Raymond, 
Cox, \& Smith [1976]).

\paragraph{Synchrotron Emission}

In a hot plasma with equipartition magnetic field, synchrotron radiation is an
important cooling mechanism.  An expression for the spectrum of optically 
thin synchrotron emission from a relativistic Maxwellian distribution of 
electrons was derived by  Mahadevan, Narayan, \& Yi (1996).  
However, below some critical frequency, $\nu_c$, the emission becomes 
self-absorbed and the optically thin emission formalism no longer applies.  
To estimate $\nu_c$ we equate the synchrotron emission from 
a sphere of radius $R$ to the Rayleigh-Jeans blackbody emission from the 
surface of that sphere. 
Below $\nu_c$ the emission is optically thick and we use the blackbody 
estimate, while above $\nu_c$ we use the expression derived by Mahadevan 
et al. (1996).  Integration over frequency yields the
total cooling rate per unit volume (Esin et al. 1996).

\paragraph{Inverse Compton Scattering}
Both bremsstrahlung and synchrotron emission mechanisms produce mainly soft 
photons, which are then upscattered by the hot electrons in the accreting gas.
The optical depth to electron scattering is simply $\tau_e = 2 n_e \sigma_T H$,
where $H$ is the scale height of the flow determined through the conditions
for vertical hydrostatic equilibrium.
After each scattering, the energy of a photon is increased by a factor 
$A = 1+4 \theta_e + 16 \theta_e^2$ (Rybicki \& Lightman 1979), where $\theta_e
= k T_e/m_e c^2$ is the dimensionless electron temperature.  The average
ratio of the photon energy at escape to its initial energy is given by
(Esin et al. 1996)
\begin{equation}
\label{eta}
\eta = e^{(A-1) s} [1-P (j_m + 1)] + \eta_{max} P (j_m+1,s),
\end{equation}
where $P (a,x) = [1/\Gamma(a)] \int^x_0 {t^{a-1} e^{-t} d t}$ is the 
incomplete gamma function and 
\begin{eqnarray}
s &=& \tau_e + \tau_e^2, \\
j_m &=& \ln{(\eta_{max})}/\ln{(A)}, \\
\eta_{max} &=& 3 k T_e/h \nu.
\end{eqnarray}

The synchrotron spectrum is strongly peaked at $\nu_c$, so we can estimate the
Compton cooling rate due to the scattering of synchrotron photons as 
$q^-_{C,s} = (\eta (\nu_c)-1) q^-_{sync}$.
On the other hand, the spectrum of bremsstrahlung radiation is practically 
flat between the synchrotron self-absorption frequency $\nu_c$ (we found that 
the bremsstrahlung self-absorption frequency is generally lower than $\nu_c$
for the range of accretion flow parameters that is relevant for this work) and 
exponential cut-off at $kT/h$.  Below $\nu_c$ free-free emission falls off
as Rayleigh-Jeans blackbody and Comptonization is not important,
since the photons are more likely to be absorbed than scattered.  Consequently,
we can estimate the emission per unit 
frequency as $q^-_{br, \nu} \simeq q^-_{br}/\left(\frac{k T}{h} - 
\nu_{c}\right)$.  Then the Compton cooling rate is given by the integral 
$q^-_{C,br} = \int_{\nu_{c}}^{\frac{k T}{h}}{q^-_{br, \nu} [\eta (\nu) -1]
d \nu}$.

The total Compton cooling rate is simply $q^-_C = q^-_{C,s} +q^-_{C,br}$.

\subsubsection{Energy Balance}

The local energy balance in the gas requires that a fraction $(1-f)$ of the
viscously dissipated energy is radiated away.  This gives us the condition
\begin{equation}
(1-f) q^+_v = q^-, 
\end{equation}
which together with the equation of state and self-similar solution described
in \S2.1 is sufficient to obtain a closed set, necessary to solve for all 
one-temperature flow parameters.  
For a two-temperature solution, we need another equation to determine $T_i$
and $T_e$ separately.  We obtain it by requiring that the net cooling rate 
has to be equal to the heating rate of the electrons:
\begin{equation}
q^- = q_{ie}.
\end{equation}

\subsection{Physical Properties of One- and Two-Temperature Flows}

The set of equations described in \S2.1-2.2 allows all the major parameters
describing the 
accreting gas to be solved for self-consistently for given values of
$M,\ \dot{M},\ R,\ \alpha$, and $\beta$.  The resulting flows are
quasispherical, with $H \sim R$, optically thin, and quite hot.  Their 
properties were described in detail by Narayan \& Yi (1995) and Esin et al. 
(1996), and here we give a brief summary.

As was shown first by Rees et al. (1982) 
advection-dominated accretion solutions are limited to 
values of $\dot{m}$ that lie below a maximum mass accretion rate, 
$\dot{m}_{crit} (r)$.  Above this limiting value, the radiative efficiency of 
the gas is so high due to the increased plasma density, that the flow cools 
down to the standard thin accretion
disk configuration.  Figure 1(a) shows a comparison of
the critical accretion rate as a function of $R$ computed for one- and 
two-temperature models.  At large radii, $r > 10^3$, the two 
calculations give identical results, with $\dot{m}_{crit} \propto r^{-3/2}$. 
Note that previous papers by Narayan \& Yi (1995) and Abramowicz et al. 
(1995) have $\dot{m}_{crit} \propto r^{-1/2}$ in the outer regions. 
Our much stronger dependance of the critical accretion rate on radius 
is due to including line cooling, which is significantly more efficient 
than bremsstrahlung at low temperatures.
Closer to the accreting object, $\dot{m}_{crit}$ decreases steeply as 
$\dot{m}_{crit} \propto r^{3/2}$ in a one-temperature model, but is roughly 
independent of $r$ in a two-temperature model (the slight decrease in 
$\dot{m}_{crit}$ seen on the figure is most likely an artifact of the 
self-similar assumption we used to derive the solution; when proper boundary 
conditions are used at the inner edge of the disk, this decrease is no
longer present [c.f. Narayan 1996]).

To understand why the two solutions behave in this way, we show 
in Figure 1(b) the radial temperature profile of the accreting gas 
for a $10 M_{\odot}$ black hole, with $\dot{M} = 10^{-5} \dot{M}_{Edd}$, 
$\alpha = 0.1$ and $\beta = 0.5$.   At large radii, the ion and electron 
temperature in a two-temperature flow are equal to each other and to the
gas temperature in a one-temperature flow.  In that region, the Coulomb 
energy transfer from ions to electrons is efficient and the two solutions
are identical.  However, in the region inside $1000 R_{Schw}$, the electron 
temperature in a two-temperature solution remains roughly constant, whereas 
the ions stay at near virial temperatures with $T_i \propto r^{-1}$.   The
gas in a one-temperature flow, on the other hand, necessarily finds an 
equilibrium temperature with $T_e = T_i = T_0$. The temperature $T_0$ 
obviously lies between $T_e$ and $T_i$ of the two-temperature solution.
At $\sim 10^{10}\ {\rm K}$, the rate of synchrotron and Compton cooling of 
the gas is very sensitive to the temperature of the electrons, and since 
electrons in one-temperature solutions are hotter than electrons in 
two-temperature solutions, the former has a much higher radiative efficiency.  
This in turn leads to lower $\dot{m}_{crit}$.

The difference in temperatures between the two solutions also leads to 
different values of the advection parameter $f$.  Figures 1(c) and 1(d) show
contours of constant $f$ as a function of radius and accretion rate, for
two- and one-temperature models respectively.
Two-temperature flows are advection-dominated, $f \sim 1$ everywhere,
except for a narrow region near the $\dot{M}_{crit}$ boundary 
(shown as a heavy line).  On the other hand, the inner regions of 
single-temperature accretion disks are cooling-dominated, $f\lsim 0.5$, for
$\dot{m} \gsim 10^{-5} \alpha^2$, because they have higher electron 
temperatures.  In fact, one-temperature hot accretion flows at high 
$\dot{M}$ have radiative efficiency 
close to $15\%$, which means they radiate away nearly all the gravitational 
energy lost in the accretion process (Esin et al. 1995).  By comparison, the 
efficiency of two-temperature accretion solutions can be orders of 
magnitude less.

\section{Effects of Nonlocal Radiation}

The formalism described in the previous section ignores the effects of 
non-local radiative transfer.  It assumes that the photons emitted at one 
radius interact only with the `local' gas, i.e. gas within a region of unit 
radial width in $\log{(r)}$.  We can check the validity of this assumption by 
estimating the optical depth for electron scattering in the radial direction 
predicted by advection models. 
The optical depth from radius $r_{min}$ to $r$ is simply $\tau_r (r_{min},r) = 
\int_{r_{min}}^r {n_e \sigma_T R_{Schw} d r}$.  After substituting the 
expression 
for $n_e$ given by Eq. (\ref{ne}), where we take $c_3 \simeq 0.3$ 
(see Appendix B), and integrating, we obtain 
\begin{equation}
\label{tau}
\tau_r (r_{min},r) = \tau_r (r_{min}) \simeq 1.8 
\left(\frac{\dot{m}}{10^{-2}}\right) 
\left(\frac{0.1}{\alpha}\right) \left(\frac{3}{r_{min}}\right)^{1/2},
\end{equation}  
assuming that $r >> r_{min}$.  Thus, two-temperature accretion disks are 
optically thin if $\dot{M} \lsim 3 \times 10^{-2} \alpha
\dot{M}_{Edd}$, which means that
the photons can travel freely through the disk and be scattered by electrons 
far from where they were emitted, heating or cooling the gas as
a result of scattering. 

One-temperature accretion disks are geometrically thinner in the inner region 
than two-temperature disks, because they are able to cool more efficiently 
(Esin et al. 1996).  As a
result, both $n_e$ and $\tau_r$ are higher.  However, even one-temperature 
flows are optically transparent below $\dot{M} \sim 3 \times 10^{-3} \alpha 
\dot{M}_{Edd}$.
 
In the remainder of this section we attempt to quantify how Compton 
scattering of non-local 
radiation will affect the flow parameters calculated based on local analysis.  
In each case 
we estimate analytically the region in the parameter space where non-local 
effects can be important, then we compute numerically how the properties 
of the solutions are affected in this region.

\subsection{Photons from the Hot Inner Region}

Both one- and two-temperature accretion flows produce most of their emission 
from within $100 R_{Schw}$, where most of the 
gravitational potential energy is released. The inner region also has 
the highest electron temperatures so that the emerging photons are hotter
than the gas in the surrounding flow.  When this radiation interacts 
with the gas in the outer regions of the disk, two processes can take place.  
Where the gas is relatively hot, less energetic photons will cool the
electrons through inverse Compton scattering.  On the other hand, where the 
temperature of the electrons is less than the effective temperature of the 
radiation, the electrons will be heated via Compton scattering. 

The amount of the heating or cooling can be easily estimated if we assume 
that each photon undergoes at most one scattering at any given radius (a 
reasonable assumption, since the Compton $y$-parameter (Rybicki \& Lightman 
1979) of the accreting gas 
at $R>100 R_{Schw}$ is always less than unity).  With this approximation, 
the rate of energy transfer from photons to electrons per unit volume, via 
Compton scattering, is just
\begin{equation}
\label{en_tr}
q_{nl} = \sigma_T c n_e e^{-\tau_r} \int{n_{\gamma,\nu}\,\Delta E_{\nu}\,
d \nu}\ {\rm erg\,s^{-1} cm^{-3}},
\end{equation}
where $n_{\gamma, \nu}$ is the number density of the incident photons per 
unit frequency and $\Delta E_{\nu} = E_{\gamma,{\rm in}} - 
E_{\gamma,{\rm fin}}$ is the energy lost by the photon in each scattering.  
When $q_{nl}$ is positive, the electrons are heated by the incident 
photon flux, when $q_{nl}$ is negative, the electrons are cooled.  
We approximate $n_{\gamma, \nu}$ as $n_{\gamma, \nu} = F_{\nu}/c h \nu  
\simeq L_{\nu}/4 \pi R^2 c h \nu$, where $L_{\nu}$ is the luminosity per 
unit frequency emerging from the region within $100 R_{Schw}$.  
$\Delta E_{\nu}$ is given by Rybicki \& Lightman (1979)
\begin{equation}
\label{deltaE}
\Delta E_{\nu} = E_{\gamma,{\rm in}} - E_{\gamma,{\rm fin}} \simeq 
\frac{h \nu}{m_e c^2} (h \nu - 4 k T_e).
\end{equation}
Since $\Delta E_{\nu}$ can be at most as large as the original photon energy,
Eq. (\ref{deltaE}) is valid only for nonrelativistic photons with 
$h \nu < m_e c^2$.  To extend this formula to higher photon energies, we 
have to replace $h \nu/m_e c^2$ by $h \nu/(m_e c^2-h \nu)$, to ensure that 
$\Delta E_{\nu} < h \nu$.  However, numerical calculations of $q_{nl}$
(described in \S3.1.1 and \S3.1.2) show that this modification has a very 
small 
effect on the final result and so for simplicity we use Eq. (\ref{deltaE}) in
our analytic estimates.

\subsubsection{Heating of the Gas}

Let us consider first the rate of heating of the gas by the energetic photons. 
From Eq. (\ref{deltaE}) it is clear that only photons 
with energies $h \nu > 4 k T_e(r)$ can contribute to the heating, whereas 
photons with lower energies will cool the gas.  The high energy cutoff of the
spectrum is determined by the electron temperature in the inner regions, 
$T_{e,in}$, since
inverse Compton scattering cannot produce photons with energies higher than
$\sim 3 k T_{e,in}$ (Rybicki \& Lightman 1979).  Therefore, to obtain an upper 
limit on the amount of heating, the integral in Eq. (\ref{en_tr}) has to be
evaluated from $\nu = 4 k T_e(r)/h$ to $\nu = 3 k T_{e,in}/h$.  Clearly then,
significant amount of Compton heating can occur only if  $T_e(r) \ll 
T_{e,in}$, which restricts us to the region outside $1000 R_{Schw}$ 
(see Figure 1[b]). 

The precise shape of $L_{\nu}$ depends sensitively on the value of the 
accretion rate.  However, a reasonable upper limit to the effect of heating 
can be obtained by assuming
that the spectrum is flat, i.e. equal energy is emitted per logarithmic 
interval in frequency, and that the luminosity emitted from the inner region
is approximately equal to the total luminosity of the disk.  Since the 
spectra generally span $\sim 10$ orders of magnitude in frequency (e.g. 
Narayan, Yi, \& Mahadevan 1995; Lasota et al. 1996), 
we can write $L_{\nu} \simeq L/(10\,{\rm log}_e 10\,\nu)$,
with $L$ being the total emitted luminosity.  Using this approximation, and
substituting Eqs. (\ref{ne}) and (\ref{deltaE}) into Eq. (\ref{en_tr}),
we obtain an approximate expression for the rate of heating caused by 
non-local radiation from smaller radii:
\begin{equation}
\label{heat1}
q^+_{nl} \simeq 1.1 \times 10^{-18}\frac{\dot{m} e^{-\tau_r} 
L}{m^3 \alpha r^{7/2}} \left(\frac{T_{e,in}}{10^9\,{\rm K}}\right)\ 
{\rm erg\,s^{-1} cm^{-3}}.
\end{equation}
where we assumed $c_3 \simeq 0.3$ (Appendix B) and used the fact that 
$T_{e,in} \gg T_e$ to simplify the final result.

%Combining Eqs. (\ref{Lnu2}) and (\ref{heat1}) gives us the final expression
%\begin{equation}
%\label{heat2}
%q^+_{nl} \simeq \left\{\begin{array}{ll}
%3.3\times 10^{20} \left(\frac{\dot{m}^3}{m^2 \alpha^3}\right)
%\left(\frac{e^{-\tau_r}}{r^{7/2}}\right) \left(\frac{\beta}{0.5}\right)
%\left(\frac{T_{e,in}}{10^9\,{\rm K}}\right)\ {\rm erg\,s^{-1} cm^{-3}} &
%\dot{m} \gsim 10^{-3} \alpha^2, \\
%6.3\times 10^{16} \left(\frac{\dot{m}^2}{m^2 \alpha}\right)
%\left(\frac{e^{-\tau_r}}{r^{7/2}}\right) \left(\frac{1-\beta}{0.5}\right)
%\left(\frac{T_{e,in}}{10^9\,{\rm K}}\right)\ {\rm erg\,s^{-1} cm^{-3}} &
%\dot{m} \lsim 10^{-3} \alpha^2.
%                            \end{array} \right.
%\end{equation}

Compton heating of electrons by hot photons becomes important when it is 
comparable to the rate of energy transfer from ions to electrons,
so to find the region of the parameter space where non-local heating must be
taken into account, we impose the condition $q^+_{nl}/q_{ie} = 
q^+_{nl}/q^- \geq 1$.
The expression for $q^-$ is derived in Appendix B, and the left-hand side
is given by Eq. (\ref{heat1}).  

For two-temperature accretion disks, this condition requires that
\begin{equation}
\label{rmdot}
\dot{m} \gsim 2.3 \times 10^{-2} \left(\frac{\alpha}{0.1}\right)
\left(\frac{1000}{r}\right)^{1/2} 
\left(\frac{10^9\,{\rm K}}{T_{e,in}}\right) e^{\tau_r},
\end{equation}
for $r > 1000$.  The values of $\dot{m}$ and $r$ for which this relation is 
satisfied lie above the dashed line on Figure 2(a), computed 
assuming $\alpha = 0.3$, $\beta = 0.5$ and $\tau_r \ll 1$.  One can see that 
the region where significant Compton heating is expected to occur is small.  

In reality, since we used upper limits in  estimating $q^+_{nl}$, our 
analysis overestimates the magnitude of this effect.  In order to find the 
exact values,  we have solved the complete set of equations discussed in 
\S2 numerically, and computed the values of electron density, plasma 
temperatures, and magnetic field strength in a two-temperature accretion 
flow for various values of $r$ and $\dot{m}$.   Using these parameters we 
calculate the local cooling rate, $q^-$, and the shape of the spectrum of 
radiation emerging from the inner region (see Narayan, Barret \& McClintock 
[1997] for details).   Then the integral in 
Eq. (\ref{en_tr}) can be computed numerically using exact values for $n_e$, 
$L_{\nu}\,e^{-\tau_r}$, and $\Delta E$, to obtain $q^+_{nl}$.
On Figure 2(a) we have plotted the contours of constant $q^+_{nl}/q^-$ as 
dotted lines.  It is clear from this figure that non-local radiation 
mainly Compton {\it cools} the accreting gas, since the values of this ratio
are for the most part negative.  Where Compton heating does occur, it 
makes up at most a few percent of the energy transferred to electrons from 
ions, and can therefore be completely neglected. 

For one-temperature flows, the total disk luminosity can be estimated from 
Esin et al. (1996) 
\begin{equation}
\label{L1}
L_{1-t} \simeq \left\{\begin{array}{ll}
1.3 \times 10^{38}\,m \dot{m}\ {\rm erg\,s^{-1}} & 
                  {\rm for}\ \ \dot{m} \gsim 10^{-5} \alpha^2, \\
1.3 \times 10^{43} \left(\frac{m \dot{m}^2}{\alpha^2}\right) 
\left(\frac{0.5}{\beta}\right)\ {\rm erg\,s^{-1}} &
                  {\rm for}\ \ \dot{m} \lsim 10^{-5} \alpha^2.
		       \end{array} \right.
\end{equation}
Then the condition $q^+_{nl} \geq q^-$ requires that 
\begin{equation}
\label{mdot2}
\dot{m} \gsim 4.6 \times 10^{-5}\,\left(\frac{\alpha}{0.1}\right)
 \left(\frac{1-\beta}{0.5}\right) \left(\frac{10^3}{r}\right)^{1/2}
 \left(\frac{10^{10}\,{\rm K}}{T_{e,in}}\right) e^{\tau_r},
\end{equation}
for $r > 1000$.  The region where Eq. (\ref{mdot2}) is valid 
lies above the dashed line on Figure 2(b), for $\alpha = 0.1$,
$\beta = 0.5$, and $\tau_r \ll 1$.  Note that the area affected by non-local 
heating is larger than the similar estimate for a two-temperature flow
on Figure 2(a).  This happens because one-temperature disks are more luminous 
than their 
two-temperature counterparts.  However, this result is again an overestimate.
As we did in the case of a two-temperature flow, we computed the ratio 
$q+_{nl}/q^-$ numerically, solving the flow equations appropriate for a
one-temperature plasma.  The constant value contours of this ratio are 
shown on Figure 2(b) (dotted lines).  The results show that Compton heating 
is indeed stronger here than it was in a two-temperature gas.  It is most 
important
for values of the mass accretion rate $\dot{m} \sim \dot{m}_{crit} (r =3)$, 
since at higher $\dot{m}$ the hot flow can not extend all the way to 
$3 R_{Schw}$, and at lower $\dot{m}$ the optical depth is low and 
Comptonization is inefficient, so that fewer high energy photons are 
produced, capable of heating the gas in the outer flow.  However even at its
maximum, non-local Compton heating does not exceed $\sim 10\%$ of the local 
heating of electrons via viscous dissipation.  

We conclude that Compton heating can be neglected within the 
advection-dominated  zone both for two- and one-temperature flows.

\subsubsection{Cooling of the Gas}

The photons with energies $h \nu < 4 k T_e$ cool the gas through inverse 
Compton scattering.  An upper limit on the amount of cooling induced by non
local radiation can be obtained by assuming that all photons are emitted
at $\nu \ll 4 k T_e/h$, so that Eq. (\ref{deltaE}) reduces to 
$\Delta E \simeq  -h \nu (4 \theta_e +16 \theta_e^2) = h \nu (1 - A)$, 
where the term $16 \theta_e^2$ is added to account 
for relativistic effects at $\theta_e \gsim 1$.  Since at $r > 100$ the
electron temperature in both one- and two-temperature flows varies as 
$T_e \simeq 10^{12} \beta\,{\rm K}/r$ (see Figure 1(a) and Eq. [\ref{temp}]), 
the quantity $A-1$ can be written as $A-1 \simeq 670 \beta/r + 
(670 \beta/r)^2$.  With this simplification, the integral in 
Eq. (\ref{en_tr}) is trivial and the expression for the cooling rate 
evaluates to 
\begin{equation}
\label{cool}
q^-_{nl} = 5.0 \times 10^{-17}\,\left[\frac{670 \beta}{r} +
\left(\frac{670 \beta}{r}\right)^2\right]\,
\frac{\dot{m} e^{-\tau_r}L} {m^3 \alpha r^{7/2}}\ {\rm erg\,s^{-1}\,cm^{-3}},
\end{equation}
where we have set $q^-_{nl} = |q_{nl}|$.

Substituting Eq. (\ref{L2}) for $L$ and imposing the requirement 
$q^-_{nl}/q^- \geq 1$, we find that in two-temperature flows 
photons from the
inner region will contribute to the cooling of the gas at $r>100$ when
\begin{equation}
\label{mdot3}
\dot{m}\gsim 6 \times 10^{-4} \left(\frac{\alpha}{0.1}\right)
\left(\frac{r}{100}\right)^{1/2} \left(\frac{0.5}{\beta}\right)
\left[1+3.4\left(\frac{100}{r}\right) 
\left(\frac{\beta}{0.5}\right)\right]^{-1} e^{\tau_r}.
\end{equation}
For $\alpha = 0.3$, $\beta= 0.5$, and $\tau_r \ll 1$, this inequality defines 
a region that lies above the dashed curve on Figure 2(c).  We see that the 
extra cooling  can be important only for high values of the accretion rate, 
when the electron density in the flow is large enough to scatter the outgoing
photons from smaller radii. 

The results of an exact numerical calculation are also shown on Figure 2(c).
We have computed $q^-_{nl}$ and $q^-$ as described in \S3.1.1, and plotted 
the curve with $q^-_{nl}/q^- = 1$ as a solid line.
The shaded region inside this curve, where non-local cooling can affect the 
accretion flow, is very small.  To investigate how the flow properties change 
in the affected region, we have included $q^-_{nl}$ into our numerical 
calculations as an extra cooling mechanism, replacing Eq. (\ref{q-}) by
$q^- = q^-_{br} + q^-_{sync} + q^-_{C} + q^-_{nl}$.  As a result, 
the equilibrium electron 
temperature of the accreting gas and the advection parameter $f$ decrease 
slightly, to compensate for the increased cooling.  However, since most of the
emission produced by the flow comes from within $100 R_{Schw}$, there is 
practically no change in the observed spectrum.

To estimate where non-local cooling is significant in one-temperature 
flows, we substitute Eq. (\ref{L1}) into Eq. (\ref{cool}) and compare the 
result to the local cooling rate, $q^-$, obtaining the following 
condition on $\dot{m}$ and $r$
\begin{equation}
\label{mdot4}
\dot{m} \gsim 1.2\times 10^{-7} \left(\frac{\alpha}{0.1}\right)
\left(\frac{1-\beta}{\beta}\right) \left(\frac{r}{100}\right)^{1/2} 
\left[1+3.4\left(\frac{100}{r}\right) 
\left(\frac{\beta}{0.5}\right)\right]^{-1} e^{\tau_r},
\end{equation}
for $r>100$.   The region of the parameter space where this inequality is 
satisfied lies above the dashed line on Figure 2(d), computed for $\alpha = 
0.1$, $\beta = 0.5$, and $\tau_r \ll 1$.   The results of a detailed numerical
computation are shown as a shaded region on the same figure.  

From comparison of Figures 2(c) and 2(d), it is clear that non-local cooling 
is much more important in one-temperature than in two-temperature 
accretion flows.  However, in both cases, the only effect is to decrease 
the gas temperature and the advection parameter in the affected regions of the 
flow, by a modest amount.  However, these regions do not produce most of the 
observed emission.  Therefore, we conclude that the radiation emitted in the 
inner regions of a hot flow does not have significant observable effect on 
the gas in the outer regions. 

\subsection{Cold Photons from the Inner Thin Disk}

In a one-temperature accretion flow model, the value of the critical
accretion rate, $\dot{m}_{crit}$, increases as $\propto r^{3/2}$ for $r \lsim 
1000$ (see Fig. 1[a]).  For $\dot{m} > \dot{m}_{crit}$, the only equilibrium 
solution is the thin accretion disk. As a result, when the value of the 
accretion rate is in the range $10^{-3} \alpha^{3/2} \lsim
\dot{m} \lsim \alpha^{3/2}$, the only possible flow configuration is a thin
disk surrounded by a hot quasi-spherical flow.  Cold photons emitted by 
the thin disk will Compton cool the hot gas around it. 
This is a similar problem to the one we discussed in \S3.1.2.  However, since
the luminosity considered here is higher because of the higher efficiency of 
the thin disk, we expect that non-local Compton cooling will be more important.

In this scenario, non-local cooling is no longer limited to the outer regions
of the disk, where the accreting gas has low density and temperature and 
therefore, multiple scatterings of the incident photons are not important.
In fact, when we compute the Compton $y$-parameter we find that it can be 
much greater than unity for $r \ll 100$.  Thus, to estimate the rate of 
non-local Compton cooling, we need to modifying the expression for 
$\Delta E_{\nu}$ to include the effect of multiple 
scatterings: 
\begin{equation}
\label{deltaE2}
\Delta E_{\nu} = E_{\gamma,{\rm in}} - E_{\gamma,{\rm fin}} = 
h \nu(1 - \eta (\nu) A),
\end{equation}
with $\eta$ given by Eq. (\ref{eta}). In general, the value of $\eta$ depends
on the frequency of the scattered radiation, but for simplicity we 
set it equal to some average value $\tilde{\eta}$.  As in \S3.1.2, we assume 
that the radiation is cold and the factor $h \nu$ is small as compared to 
$k T_e$.   With that approximation, the Compton cooling rate becomes 
\begin{equation}
\label{thin_in}
q^-_{nl} = \sigma_T n_e e^{-\tau_r} F_c (\tilde{\eta}A-1).
\end{equation}
where $F_c = L_c/4 \pi R^2$ is the cold photon flux and $L_c$ is the total 
luminosity of the thin disk.  Since the standard thin disk has
approximately Keplerian energetics, we can estimate $L_c$ as the energy loss 
between the outer and inner radii of the disk: 
\begin{equation}
\label{lcold}
L_c \simeq \frac{G M \dot{M}}{2 R_{Schw}} \left(\frac{1}{3} - 
\frac{1}{r_{out}}\right).
\end{equation} 

Combining Eqs. (\ref{thin_in}) and (\ref{lcold}), we obtain the following
expression for the cooling rate
\begin{equation}
\label{thin_in2}
q^-_{nl} \simeq 1.5 \times 10^{23} \left(\frac{\dot{m}^2}
{m^2 \alpha}\right) \left(\frac{e^{-\tau_r}}{r^{7/2}}\right) 
(\tilde{\eta}A-1) \left(\frac{1}{3} - 
\frac{1}{r_{out}}\right).
\end{equation}
Compton cooling becomes important when it exceeds local cooling of the gas. 
Comparing Eqs. (\ref{thin_in2}) and (\ref{q-loc}) we find that $q^-_{nl}$
is greater than $q^-$ at all radii, provided that $\dot{m} \gsim 10^{-3} 
\alpha^2$ (necessary for the thin disk to be present) and 
$\tilde{\eta}A-1 > 1$.   Thus, if the thin disk 
is present, it will always Compton cool the hot flow around it.  

To obtain more detailed results, we computed the quantities $q^-$ and 
$q^-_{nl}$ numerically.  In our calculation, we assumed that all the 
radiation from the thin disk is emitted at the frequency corresponding to 
a blackbody of temperature $T_{BB} \simeq (3 G M 
\dot{M}/8 \pi R^3 \sigma_B)^{1/4}$ 
(Frank, King, \& Raine 1992) evaluated at $R=3 R_{Schw}$, and computed the 
corresponding $\tilde{\eta}$ at each radius.  The region where $q^-_{nl}/q^- 
\ge 1$ is shaded on Figure 3(a).  The flow in this region will 
be cooled significantly by photons from the thin disk. 
Including non-local Compton cooling directly in our calculations shows that 
because of the increase in the cooling rate, the gas near the inner edge of 
the hot flow collapses to a thin disk, and the critical accretion rate line 
itself moves downwards.  This means that the thin disk 
extends slightly further outward and produces more cold radiation.  
After a few iterations, this process converges;  the new
$\dot{M}_{crit}$ is shown on Figure 3(a) as a dashed line.  Due to 
extra cooling, the maximum accretion rate for which hot one-temperature 
solution exists, decreased by a factor of $\sim 10$.
For $\dot{M} < \dot{M}_{crit}$, where hot solutions are still allowed, 
the equilibrium temperature of the accreting gas decreases (Figure 3[b]), 
so that the observed spectrum contains less high energy flux.

\subsection{Cold Photons from the Outer Thin Disk}

Outside $1000 R_{Schw}$,  the critical accretion rate 
for both one- and two-temperature solutions decreases with radius as 
$\dot{m}_{crit} \propto r^{-3/2}$.  Therefore, for any value of $\dot{m}$ 
there is a critical radius $r_{crit}$ beyond which the only possible flow 
configuration is the thin disk.  Therefore, we expect to have accretion 
flows where the gas forms a cool thin disk for $r > r_{crit}$, but switches 
to a hot advection-dominated state for $r < r_{crit}$.   Many of the models 
published in the literature (e.g. Narayan et al. 1996) are of this form.

In such configurations, the photons produced in the cold outer flow  may be
expected to Compton cool the hot gas inside $r_{crit}$.  The cooling rate per 
unit volume is given by Eq. (\ref{thin_in}), the only unknown being the cold 
photon flux, $F_c$.

In general, $F_c$ is a function of both
radius, $R$, and the height above the plane of the thin disk, $z =
R \sin{\theta}$.  Taking
into account the full geometry of the problem, we write $F_c (R,z)$ as
a double integral:
\begin{equation}
\label{2int}
F_c (R,z) = \int^{\infty}_{R_{tr}}{d R^{\prime}\ 2 \int_0^{\pi}{R^{\prime}\, 
d \phi\, \frac{\sigma_B T_c^4 \sin{\psi}}{\pi x^2}}}, 
\end{equation}
where $R_{tr}$ is the transition radius between the hot and cold parts of the 
accretion disk, $T_c (R^{\prime})$ is the surface temperature of the thin disk 
at radial position $R^{\prime}$, $x$ is the distance from the emitting element 
in the thin disk to the scattering element in the hot flow, $\phi$ is the 
azimuthal angle, and $\sin{\psi} = z/x$.  Since we are working with a 
height-averaged set of equations, we replace $F_c (R,z)$ with its average 
over a spherical shell of radius $R$:
\begin{equation}
\label{3int}
\widetilde{F}_c (R) = \int^{\pi/2}_0 {\cos{\theta}\, d \theta\,
\int_{R_{tr}}^{\infty}{d R^{\prime}\ 2 \int_0^{\pi}{R^{\prime}\, d \phi\, 
\frac{\sigma_B T_c^4 \sin{\psi}}{\pi x^2}}}}
\end{equation}

The emission per unit area of the thin disk can be approximated as
$\sigma_B T_c^4 \simeq 3 G M \dot{M}/8 \pi R^3$ (Frank, King, \& Raine 1992), 
and substituting expressions for $x$ and $\psi$ in terms of $R$, $R^{\prime}$ 
and $\theta$ we obtain
\begin{equation}
\label{3int2}
\widetilde{F}_c (R) = \frac{3 G M \dot{M} R}{4 \pi^2} 
\int_{R_{tr}}^{\infty}{\frac{d R^{\prime}}{{R^{\prime}}^2}
\int^{\pi/2}_0 {d \theta\,\cos{\theta}\,\sin{\theta}
\int_0^{\pi}{\frac{d \phi}{({R^{\prime}}^2 + R^2 - 2 R^{\prime} R \cos{\theta} 
\cos{\phi})^{3/2}}}}}
\end{equation}
This integral has to be evaluated numerically, but a reasonable estimate
can be obtained by replacing $\sin{\psi}$ by $\sin{\tilde{\psi}}$, where 
$\tilde{\psi}$ is some effective angle averaged over the entire spherical 
shell.  Since
$\sin{\psi} = R \sin{\theta}/({R^{\prime}}^2 + R^2 - 2 R^{\prime} R 
\cos{\theta} \cos{\phi})^{1/2}$, setting $\phi = 0$, 
$\theta = \cos^{-1}{R/R_{tr}}$, and $R^{\prime} = R_{tr}$ gives an upper 
limit of $\sin{\psi} = R/R_{tr}$.  Then the three integrals can be evaluated 
analytically:
\begin{equation}
\label{flux}
\widetilde{F}_c (R) = \frac{3 G M \dot{M}}{4 \pi} \frac{R}{R_{tr}} \left\{ \frac{1}
{2 R_{tr} R^2} + \frac{R^2 - R_{tr}^2}{4 R_{tr}^2 R^3} \left[\ln{\left(1+
\frac{R}{R_{tr}}\right)} -
\ln{\left(1-\frac{R}{R_{tr}}\right)} \right] \right\}.
\end{equation}
In the limit when $R \sim R_{tr}$, Eq. (\ref{flux}) reduces to 
$\widetilde{F}_c \simeq 3 G M \dot{M}/8 \pi R_{tr}^2 R$; when
$R \ll R_{tr}$, $\widetilde{F}_c \simeq G M \dot{M} R/4 \pi R_{tr}^4$, so that 
when $R/R_{tr} \rightarrow 0$, the cold photon flux becomes completely 
negligible.

Clearly, Compton cooling will be strongest near the boundary between 
the hot and cold flows, so an upper limit on the amount of extra cooling can
be obtained by replacing Eq. (\ref{flux}) by the limiting expression for 
$\widetilde{F}_c$ when $R \sim R_{tr}$.  Substituting this result into Eq. 
(\ref{thin_in}) and imposing the familiar constraint $q^-_{nl}/q^- \geq 1$, 
we find that external cooling is significant in the hot flow near the 
boundary if
\begin{equation}
\label{r_tr1}
r_{tr} \lsim 200 \left(\frac{r}{r_{tr}}\right)^3 
\left(\frac{\alpha}{0.1}\right)^2 \left(\frac{0.5}{\beta}\right)^2 
\left(\frac{\tilde{\eta} A(r) -1}{2}\right)^2,
\end{equation}
for $\dot{m} \gsim 10^{-3} \alpha^2$, and
\begin{equation}
\label{r_tr2}
r_{tr} \lsim 10^3 \left(\frac{r}{r_{tr}}\right)^3 
\left(\frac{\dot{m}}{10^{-5}}\right)^2 \left(\frac{0.5}{1-\beta}\right)^2
\left(\frac{0.1}{\alpha}\right)^2 (\tilde{\eta} A(r)-1)^2,
\end{equation}
for $\dot{m} \lsim 10^{-3} \alpha^2$.  Since the critical outer radius for a
hot flow is always greater than $10^3 R_{Schw}$ (see Figure 1[b]), this result
shows that the thin disk radiation has no effect on the hot gas, so long as  
$R_{tr} \sim R_{crit}$.

The cooling from the outer disk can, however, be significant if $R_{tr} \ll 
R_{crit}$.  There is no well-developed theory at this time for the transition 
radius $R_{tr}$ and one could in principle imagine a situation where the gas 
remains in a cold state down to a much smaller radius than $R_{crit}$.  If
we consider such a model, the extra cooling of the hot gas at $R \sim R_{tr}$ 
can be quite important.   To show this, we computed the ratio $q^-_{nl}/q^-$ 
numerically, using the full 
expression for $\widetilde{F}_c$ (as given by Eq. [\ref{flux}]).  
The results for one- and two-temperature hot flows 
are shown on Figures 4(a) and 4(b) respectively, where we have plotted
$q^-_{nl}/q^-$ evaluated at $R = R_{tr}$ for different values of
$\dot{m}$ and $r_{tr}$.  We can see that external cooling begins to dominate 
when $r_{tr} \sim 10^2-10^3$, in reasonable agreement with our analytic 
approximation.  

As the transition radius decreases further, the ratio of non-local to local 
cooling at the boundary between a two-temperature hot flow and a thin disk 
becomes more extreme, increasing up to $\sim 200$ for high accretion rates,
and causing a significant decrease in the equilibrium temperature of the 
hot gas.  Note however, that only a relatively thin layer near the
boundary is affected by extra cooling.  Away from the boundary, the cold 
photon flux decreases linearly with $r$ and the ratio of non-local to 
local cooling falls 
steeply with decreasing radius (see Eqs. [\ref{thin_in}] and [\ref{q-loc}]):
\begin{equation}
\label{frac}
\frac{q^-_{nl}}{q^-} \propto \frac{r\cdot r^{-3/2}}{r^{-4}} = r^{7/2}.
\end{equation}
Because of this strong dependence on $r$, the net effect on the hot flow is 
negligible, as long as the hot gas in the boundary layer is in {\em stable 
thermal equilibrium}.  However, the hot equilibrium configuration cannot 
be maintained for an arbitrarily strong cold photon flux.  At some point, the
total cooling in the boundary layer becomes so strong, that the gas temperature
is forced significantly below the virial value, i.e. the
hot quasispherical flow collapses into a cold thin disk.  
When this occurs, the transition radius effectively moves inwards, which
means that the hot gas in the new boundary layer will experience even 
stronger external cooling, and will also inevitably collapse.
In this fashion, a runaway instability develops that causes the entire hot 
flow to cool down to the standard  Shakura \& Sunyaev disk.  

We have computed numerically the values of the transition radius at which 
this instability occurs for different mass accretion rates.  This was done 
by incorporating $q^-_{nl}$ directly into the calculations and decreasing 
$r_{tr}$ until the hot stable accretion flow  solution at $r \sim r_{tr}$ 
disappeared.  The resulting critical values of $r_{tr}$ are shown on Figure 5
as dashed lines.  We see that this instability is an important consideration 
only for large values of the mass accretion rate.  
For  $\dot{M} \lsim 0.1 \times \alpha^2 \dot{M}_{Edd}$, the thin disk has 
virtually no effect on the hot flow, irrespectively of where the transition 
between the cold and hot phases occurs.

In a single-temperature hot flow, the value of the ratio $q^-_{nl}/q^-$ 
evaluated at $r = r_{tr}$
never becomes much higher than unity (see Figure 4[b]), even for small 
values of the transition radius.   The reason for this is that at 
$r < 100$, one-temperature flows are cooling-dominated, i.e. local 
cooling is very efficient.  At $r < r_{tr}$ the cooling ratio behaves
roughly in accordance with Eq. (\ref{frac}), so that external cooling is 
completely unimportant there.  

\subsection{Irradiation of the Thin Disk by the Hot Flow}

In calculating the energy flux from the thin disk in \S3.2 and \S3.3 
we used the standard (Frank, King, \& Raine 1992) formalism and ignored
the effects of irradiation of the cold gas by the advection flow.  In reality,
most of the hot photons incident on the thin disk are reprocessed 
there, so that the net emission from the thin disk is a blackbody at a 
higher temperature than predicted by the standard theory.  To test 
whether irradiation is important, we need to compute 
the ratio of the incident energy flux to the viscous dissipation rate per 
unit area of the thin disk. 

Using cylindrical geometry for the hot flow, the incident energy per unit 
area of an inner thin disk can be written as a triple integral
\begin{equation}
\label{inc}
Q_{inc}(R^{\prime}) = \int^{\infty}_{R_{tr}} {d R\ 2 \int_0^{\pi}
{R\,d \phi \int_0^{H(R)} {d z\,\frac{q^-(R) \sin{\psi}}{4 \pi x^2}}}},
\end{equation}
where $R$, $R^{\prime}$, $z$, $x$, and $\phi$ are defined as in \S3.3.  Since
the thin disk is surrounded with a one-temperature accretion flow, for which 
$f$ is significantly below unity, we can set $q^- = q^+_v$, where $q^+_v$
is given by  Eq. (\ref{q+}).  The result can be expressed in terms
of elliptical integrals, but to get a rough estimate of $Q_{inc}(R^{\prime})$
we assume that $R^{\prime} \ll R$, i.e. we restrict our analysis to radii
well inside the transition radius. Then to zeroth order in 
$(R^{\prime}/R)$ the ratio of the incident to the 
viscously dissipated energy evaluates to
\begin{equation}
\frac{Q_{inc}(R^{\prime})}{\sigma_B T_c (R^{\prime})} = 0.01 
\left(\frac{r}{r_{tr}}\right)^3.
\end{equation}
Thus, irradiation contributes at most a few percent to the energy output of 
the thin disk.  This result was confirmed by integrating 
Eq. (\ref{inc}) numerically,
with the correct expression for $q^-$ in the case of a one-temperature flow.
We conclude that irradiation is entirely negligible in this scenario, 
because the inner thin disk subtends a small solid angle from the point of 
view of the hot flow, and most of the energy is dissipated in the thin disk.  
By comparison, in the disk and corona model 
(e.g. Haardt \& Maraschi 1991), where irradiation is known to play an important
role, the hot corona lies directly  on top of the cold gas, so that roughly 
half of all photons emitted in the corona are absorbed by the disk.

The expression for the energy incident on an outer thin disk differs from 
Eq. (\ref{inc}) only in the limits of the outermost integral; in this case we
integrate $R$ from the last stable orbit, $3 R_{Schw}$, to $R_{tr}$.  
Again, we restrict ourselves to regions away from the transition radius,
i.e. $R^{\prime} \gg R$, and assume that $q^- = q^+_v$.  Then to first order 
in $R/R^{\prime}$, the ratio
of the incident to viscously dissipated energy per unit area is independent
of $R^{\prime}$:
\begin{equation}
\label{ratio2}
\frac{Q_{inc}(R^{\prime})}{\sigma_B T_c (R^{\prime})} = 0.15 \ln{\left(
\frac{r_{tr}}{3}\right)}.
\end{equation}
Numerical calculations show that even at $R^{\prime}\sim R$, the result 
increases by at most $\sim 20\%$.  

This calculation shows that irradiation of the outer thin disk becomes 
important only at large transition radii, where, according to our results 
from \S3.3 the presence of the thin disk does not affect the hot flow.  Note 
also that large transition radii are allowed only for low accretion rates,
when the flow becomes strongly-advection dominated, and the result in 
Eq. (\ref{ratio2}) decreases by an additional factor of $(1-f)$. On the
other hand, if the transition occurs within $10^3 R_{Schw}$, irradiation 
will increase the flux from the thin disk by at most a factor of 2.   This 
has only a small effect on our calculations of the critical transition 
radius, and qualitatively our results remain the same.  

\section{Discussion}

In this paper we have explored how non-local radiation transfer affects the 
properties of hot one- and two-temperature advection-dominated solutions 
described by Esin et al. (1996), Narayan \& Yi (1995) and Abramowicz et al. 
(1995).  Since these 
solutions are optically thin, radiation emitted at one radius could in 
principle induce significant cooling or heating of the accreting gas at some 
other radius, through Compton scattering.  We considered three different 
sources of non-local radiation: a luminous inner region of a hot disk ($R < 100
R_{Schw}$), a thin disk inside a hot one-temperature gas, and a thin 
Shakura \& Sunyaev disk surrounding a hot advection-dominated flow.

We found that two-temperature solutions are not affected by the radiation
from their hot interior.  These solutions are strongly advection-dominated 
everywhere except near the outer boundary of the disk, $R_{crit}$, (see
Fig. 1[c]), and the inner part simply does not produce enough hot photons to 
induce significant heating of the outer layers.  On the other hand, 
the gas density of the flow at $R > 100 R_{Schw}$ is so small, that with
electron temperatures of $T_e \lsim 10^9\ {\rm K}$, Compton cooling by
non-local radiation can not play an important role either.  

The inner parts of one-temperature flows are cooling-dominated and 
produce more radiation than their two-temperature counterparts.  However, 
single-temperature solutions are limited to lower values of the mass
accretion rate (see Figure 1[a]), 
and since hot photons are produced mainly by Comptonization, which 
is not an efficient process at low $\dot{M}$, no heating of the outer flows
occurs.  On the other hand, we find that Compton cooling by the photons from 
the interior dominates over local cooling processes for $10^2 
R_{Schw}\lsim R \lsim 10^4R_{Schw}$ and $\dot{M} \gsim 10^{-7} \dot{M}_{Edd}$. 
However, since the bulk of the observable radiation 
comes from inside $100 R_{Schw}$, the changes in the flow properties caused
by extra cooling do not appreciably affect the overall spectrum of the system. 

Compton cooling by the radiation from the inner thin disk does
have a significant impact on one-temperature flows, since it occurs at higher
$\dot{M}$, when Compton scattering is very effective.  The extra cooling 
induced by the thin disk photons is so strong, that the maximum accretion rate
for which the one-temperature model has an equilibrium solution decreases by a 
factor of $\sim 10$ (Figure 3[a]).  

Finally, we found that the radiation from an outer thin disk does not have
any effect on a hot flow that extends beyond $\sim 10^3 R_{Schw}$.  At this
distance, the gas is too cold for significant Compton cooling 
to take place.  When the transition radius between the hot and cold gas moves 
closer to the accreting object, the non-local cooling of two-temperature 
gas near the boundary can become very large.  However, this effect is 
limited to the layers of the
hot flow very close to the transition between the hot and cold gas; away 
from this boundary, the ratio of non-local to local cooling decreases as 
$r^{7/2}$.  Therefore, external cooling does not affect the regions of the
hot gas where most of the observable radiation is produced and can be 
disregarded until it is so 
strong that it forces the hot flow near the boundary to cool to a thin disk 
configuration.  When this happens, the transition radius decreases and the hot
gas in the new boundary layer experiences even stronger external cooling,
which causes it to collapse in its turn.  The resulting runaway process
continues until the entire flow assumes the thin disk configuration.  This 
phenomenon can be important if the hot advection-dominated flows are formed 
by evaporation of the thin disk near the accreting black hole (Narayan \& Yi 
1995).  In that case,
the maximum accretion rate at which such hot flow can form is in fact less
than $\dot{M}_{crit}$ computed based on the local assumption.

\acknowledgments

I would like to thank Ramesh Narayan and Eve Ostriker for many useful comments
and discussions, and my referee for his/her friendly and helpful suggestions.  
This work was supported in part by NSF grant AST 9423209.

\vfill\eject
\begin{appendix}
\section{Modified Derivation of the Energy Equation And Adiabatic Exponent for 
the Accreting Gas}

One of the conditions that the physical parameters of the accretion flow 
must satisfy is the conservation of energy.  Following Abramowicz et al. (1988)
and Narayan \& Yi (1994) we can write this condition as 
\begin{equation}
\label{cons_en}
\rho v T \frac{d s}{d R} = f q^+,
\end{equation}
where $s$ is the entropy per unit mass of the gas and other quantities are 
defined in \S2.1.  

To derive the expression for $d s/d R$ we model the gas in the accreting flow 
as a combination of an ideal monatomic gas (a justified assumption since the 
accretion flow consists mainly of ionized hydrogen) and tangled magnetic 
fields.  Radiation pressure was 
shown to be unimportant in one-temperature flows (Esin et al. 1996) and 
since two-temperature solutions are even less luminous, the radiation pressure 
can safely be ignored.  Then the total pressure in the gas is the sum of
the thermal ideal gas pressure, and the pressure due to the magnetic field
(Eqs. [\ref{pg}] and [\ref{pm}]) and can be written as
\begin{equation}
\label{ptot}
p = p_g + p_m = \frac{\rho k T}{\mu m_u} + \frac{B^2}{24 \pi} = 
\frac{\rho k T}{\beta \mu m_u},
\end{equation}
where we have defined $T = T_i$, $\mu = \mu_i \mu_e/(\mu_e + \mu_i T_e/T_i)$
and $\beta = p_g/p$. 
In general, $\mu$ is a function of both $T_i$ and $T_e$; however, for 
simplicity we restrict ourselves to the two limiting cases when $T_i \gg T_e$
and $\mu = \mu_i$, or $T_i = T_e$ and $\mu = \mu_i \mu_e/(\mu_e + \mu_i)$.

The internal energy of the gas is the sum of the kinetic energy of the
particles (since the gas is assumed to be monatomic) and the energy stored in 
the magnetic field.  The total internal energy per unit mass is then
\begin{equation}
\label{int_en}
u = \frac{3}{2} \frac{k T}{\mu m_u} + \frac{1}{\rho} \frac{B^2}{8 \pi} =
\frac{3}{2} \frac{k T}{\mu m_u} + \frac{3 (1-\beta)}{\beta} 
\frac{k T}{\mu m_u} = \frac{(6 - 3 \beta)}{2 \beta} 
\frac{k T}{\mu m_u}
\end{equation}
In a quasistatic process, the first law of thermodynamics requires that
\begin{equation}
\label{flth}
T d s = d u + p\,d V = \left(\frac{\partial u}{\partial T}\right)_V d T +
\left(\frac{\partial u}{\partial V}\right)_T d V + p\,d V,
\end{equation}
where $V = 1/\rho$ is the volume per unit mass.  But $u$ is a function of $T$ 
only, which means that the first term vanishes,
$(\partial u/\partial V)_T = 0$.  The second term we evaluate as
\begin{equation}
\label{dudT}
\left(\frac{\partial u}{\partial T}\right)_V = \frac{d u}{d T} = 
\frac{(6 - 3 \beta)}{2 \beta} \frac{k}{\mu m_u}.  
\end{equation}
Finally, we divide both sides 
of Eq. (\ref{flth}) by  $T$ and obtain the following relation for the entropy
\begin{equation}
\label{d_ent}
d s = \frac{(6 - 3 \beta)}{2 \beta} \frac{k}{\mu m_u} \frac{d T}{T} + 
\frac{\rho k}{\beta \mu m_u} d \left(\frac{1}{\rho}\right).
\end{equation}
For an adiabatic process we set $d s=0$.  Following Clayton 
(1983, Eq. [2-121c]) we define the corresponding adiabatic exponent $\gamma$ as
\begin{equation}
\label{gamma}
\gamma = \Gamma_3 = \frac{\left(\frac{1}{\beta} \frac{k}{\mu m_u}\right)}
{\left(\frac{(6 - 3 \beta)}{2 \beta} \frac{k}{\mu m_u}\right)} + 1 =  
\frac{8 - 3 \beta}{6-3 \beta}
\end{equation}
With this definition, we integrate Eq. (\ref{d_ent}) to obtain an expression 
for the entropy of the gas up to a constant factor:
\begin{equation}
\label{entropy}
s = \frac{k}{\beta \mu m_u} \frac{1}{\gamma - 1} \ln{(c_s^2 \rho^{1-\gamma})} +
{\rm const}.
\end{equation}

Now we are in a position to write down the final form for the energy 
conservation equation.  Evaluating the derivative of $s$ with respect to $R$
and substituting it into Eq. (\ref{cons_en}), yields the same expression 
as given by Narayan \& Yi (1994)
\begin{equation}
\label{cons_en2}
\frac{\rho v}{\beta (\gamma - 1)} \frac{d c_s^2}{d R} - 
v c_s^2 \frac{d \rho}{d R} = f q^+,
\end{equation}
but with a different value for $\gamma$, namely Eq. (\ref{gamma}).

\newpage
\section{Approximate Analytic Expression for the Viscous Heating and Local 
Cooling}

We need to obtain simplified expressions for the rate of viscous energy 
dissipation and local cooling of the accreting gas, as functions of the global 
parameters, $m, \dot{m}, \alpha, \beta$ and $r$ only.  

\subsection{Two-Temperature Accretion Disks} 

In two-temperature flows the ion temperature is always nearly virial and
scales as $M/R \propto 1/r$.  The electron and ion temperatures are equal 
when $r \gsim 1000$, but $T_e$ stays nearly constant at $\sim 10^{9}\,{\rm K}$ 
in the inner region of the disk.  We find that the sum $T_i + T_e$ is well 
approximated by 
\begin{equation}
\label{temp}
T_i + T_e \simeq \frac{10^{12}\,{\rm K}}{r} \left(\frac{\beta}{0.5}\right),
\end{equation}
for all values of $\dot{m}$, $\alpha$, and $\beta$.
Substituting this expression into Eq. (\ref{T}) gives us $c_3 \simeq 
0.3$.  Inverting Eq. (\ref{c3}), we obtain an expression for 
$\epsilon^{\prime}$ in terms of $c_3$:
\begin{equation}
\label{ep2}
\epsilon^{\prime} = \frac{1}{2} \left(\frac{9 \alpha^2}{x} - \frac{x}{2}-
5\right),\ \ \ \ x = \frac{9 c_3 \alpha^2}{2}.
\end{equation}
The second term in brackets can be neglected as long as 
$\alpha \lsim 0.5$.  Then Eq. (\ref{ep2}) simplifies to $\epsilon^{\prime} 
\simeq 0.8$.  

Substituting these results into Eq. (\ref{q+v}), we find that the total 
energy per unit volume dissipated in the disk can be written as 
\begin{equation}
\label{q+}
q^+_v \simeq 8.1 \times 10^{20}\,\frac{\dot{m}}{m^2 r^4}\ 
{\rm erg\,s^{-1} cm^{-3}}.
\end{equation}

As required by the energy balance equation, the rate of local cooling is
given by $q^- = (1-f) q^+_v$.  In general, $f$ is a 
function of $r$ as well as $\dot{m}$, $\alpha$, and $\beta$,  but for 
simplicity we use the radially averaged value which can be easily estimated 
by integrating $q^-$ over the volume of the disk and comparing the result
with the total emitted luminosity, $L_{2-t}$:
\begin{equation}
\label{f}
L_{2-t} \simeq (1-f) \int_{r_{min}}^{\infty} {q^+_v\, 2 H\, 2\pi R\,dR}.
\end{equation}
We adopt the value for $L_{tot}$ estimated by Mahadevan (1997)
\begin{equation}
\label{L2}
L_{2-t} \simeq \left\{\begin{array}{ll}
   1.3 \times 10^{38} \left(\frac{m \dot{m}^2}{\alpha^2}\right) 
   \left(\frac{\beta}{0.5}\right)\ {\rm erg\,s^{-1}}, &
	          \dot{m} \gsim 10^{-3} \alpha^2, \\
   2.5 \times 10^{34}\,m \dot{m} \left(\frac{1 - \beta}{0.5}\right)\ 
   {\rm erg\,s^{-1}}, &
	          \dot{m} \lsim 10^{-3} \alpha^2;
                       \end{array} \right.
\end{equation}
Combining  Eqs. (\ref{q+}), (\ref{f}), and (\ref{L2}) and setting $r_{min} =3$,
we obtain the following expression for the local cooling
\begin{eqnarray}
\label{q-loc}
q^- &=& (1-f) q^+_v = 9.3 \times 10^{-18} \frac{L_{2-t}}{m^3 r^4} \nonumber  \\
 &=& \left\{\begin{array}{ll} 
1.2 \times 10^{21} \frac{\dot{m}^2}{m^2 \alpha^2 r^4} 
  \left(\frac{\beta}{0.5}\right)\ {\rm erg\,s^{-1} cm^{-3}}, &
  \ \ \ {\rm for}\ \ \dot{m} \gsim 10^{-3} \alpha^2, \\
2.3\times 10^{17} \frac{\dot{m}}{m^2 r^4} \left(\frac{1-\beta}{0.5}\right), &
  \ \ \ {\rm for}\ \ \dot{m} \lsim 10^{-3} \alpha^2.
	\end{array} \right.
\end{eqnarray}

\subsection{One-Temperature Accretion Disks}

Two-temperature flows become effectively single-temperature when $r > 1000$, 
since there $T_e = T_i$ (Esin et al. 1996, Fig. 1[b]).
Moreover, in the region $30 \lsim r \lsim 1000$ the gas temperature, $T$, in 
one-temperature disks is nearly as high as $T_i$ in equivalent two-temperature
solutions, and the values of the advection parameter $f$ are similar as well.  
Therefore, in the outer regions Eqs. (\ref{q+}) and (\ref{q-loc}) apply 
equally well to both models.  

Inside $\sim 30 R_{Schw}$, as one-temperature disks become thinner,
denser and cooling-dominated, both $q^+_v$ and $q^-$ increase faster with 
decreasing radius than the corresponding quantities in two-temperature disks.
In this regime, we can use Eqs. (\ref{q+}) and (\ref{q-loc}) as lower limits.
\end{appendix}

\vfill\eject
\references
\def\refpar{\hangindent=3em\hangafter=1}
\def\reference{\refpar\noindent}
\def\apj{ApJ}
\def\apjs{ApJS}
\def\mnras{MNRAS}
\def\aa{A\&A}
\def\aas{A\&AS}
\def\aj{AJ}
\def\nat{Nature}

\reference Abramowicz, M. A., Chen, X., Kato, S., Lasota, J. P., \& Regev, O.
1995, \apj, 438, L37

\reference Abramowicz, M. A., Czerny, B., Lasota, J. P., \& Szuszkiewicz, E.
1988, \apj, 332, 646

\reference Begelman, M. C. \& Chiueh, T. 1988,\apj, 332, 872

\reference Chen, X., Abramowicz, M. A., Lasota, J. P., Narayan, R., Yi, I.
1995, \apj, 443, L61

\reference Clayton, D. D. 1983, Principles of Stellar Evolution and 
Nucleosynthesis, U. of Chicago Press

\reference Esin, A. A., Narayan, R., Ostriker, E., \& Yi, I. 1996, \apj, 465,
312

\reference Frank, J., King, A., \& Raine, D. 1992, Accretion Power in
Astrophysics (Cambridge, UK: Cambridge University press)

\reference Grindlay, J. E. 1978, \apj, 221, 234

\reference Haardt, F. \& Maraschi, L. 1991, \apj, 380, L51

\reference Lasota, J. P., Abramowicz, M. A., Chen, X., Krolik, J., Narayan, R.,
\& Yi, I. 1996, \apj, 462, 142L

\reference Mahadevan, R. 1997, \apj, 477, 000

\reference Mahadevan, R., Narayan, R., \& Yi, I. 1996, \apj, 465, 327

\reference Narayan, R. 1996, \apj, 462, 136

\reference Narayan, R., Barret, D., \& McClintock, J. E. 1997, submitted to 
\apj

\reference Narayan, R., McClintock, J. E., \& Yi, I. 1996, \apj, 457, 821

\reference Narayan, R. \& Yi, I. 1994, \apj, 428, L13

\reference Narayan, R. \& Yi, I. 1995, \apj, 452, 710

\reference Narayan, R., Yi, I., \& Mahadevan, R. 1995, \nat, 374, 623

\reference Ostriker, J. P., McCray, R., Weaver, R., \& Yahil, A. 1976, 
\apj, 208, L61

\reference Pacholczyk, A. G. 1970, Radio Astrophysics (San Francisco: Freeman)

\reference Phinney, E. S. 1981, in Plasma Astrophysics, eds. T. D. Guyenne \&
G. Levy (ESA SP-161), 337

\reference Rees, M. J., Begelman, M. C., Blandford, R. D., Phinney, E. S. 1982,
\nat, 295, 17

\reference Raymond, J. C., Cox, D. P., \& Smith, B. W. 1976, \apj, 204, 290

\reference Rybicki, G. B. \& Lightman, A. P. 1979, Radiative Processes in
Astrophysics (New York: John Wiley \& Sons)

\reference Shakura, N. I. \& Sunyaev, R. A. 1973, \aa, 24, 337

\reference Shapiro, S. L., Lightman, A. P., \& Eardley, D. M. 1976, \apj, 204,
187

\reference Shvartsman, V. F. 1971, Soviet Astr.-AJ, 14, 662

\reference Stepney, S., \& Guilbert, P. W. 1983, \mnras, 204, 1269

\reference Svensson, R. 1982, \apj, 258, 335

\vfill\eject
\noindent
{\bf Figure Captions}
\\

\noindent
Figure 1. (a) Critical mass accretion rate, $\dot{M}_{crit}$, as a function 
of radius for one-temperature (dashed line) and two-temperature (solid line) 
models, plotted for two different values of the viscosity parameter $\alpha$.
Hot advection-dominated solutions exist below the respective lines; above the 
lines the only equilibrium 
configuration available to the accretion flow is a standard thin disk.
(b) Radial temperature profile for a flow with a fixed accretion rate of
$10^{-5} M_{Edd}$.  
(c) Contours of constant advection parameter $f$ for a two-temperature 
accretion flow.  The critical accretion rate is shown as a heavy line.    
(d) Contours of constant advection parameter $f$ for a one-temperature 
accretion flow.  The critical accretion rate is shown as a heavy line.    \\

\noindent
Figure 2. Critical mass accretion rate (heavy line), $\dot{M}_{crit}$,
plotted in Eddington units, as a function of radius for two-temperature 
(a, c) and one-temperature (b, d) models.  
(a) and (b) Above the dashed line lies the region where significant Compton 
heating by the photons emitted at $R < 100 R_{Schw}$ is expected to occur, 
based on an analytic approximation (Eqs. [\ref{rmdot}] and
[\ref{mdot2}]).
The dotted lines are the contours of constant $q_{nl}/q^-$ computed
numerically, labeled by
the values of this ratio.  Positive numbers correspond to a net heating, 
negative numbers indicate that non-local radiation
produces a net cooling.  Note that the ratio of non-local to local heating 
is always significantly below unity.
(c) and (d) Shaded area corresponds to the region of the accretion flow 
where local cooling processes are less efficient than Compton cooling by the 
photons emitted at $R < 100 R_{Schw}$, computed numerically.  The dashed 
lines are the contours of 
$q^-_{nl}/q^-=1$, estimated analytically (Eqs. [\ref{mdot3}] and 
[\ref{mdot4}]). \\

\noindent
Figure 3. (a) Critical mass accretion rate, $\dot{M}_{crit}$,
plotted in Eddington units (solid heavy line), for a one-temperature model
with a $10 M_{\odot}$ black hole, viscosity parameter $\alpha=0.1$,
and ratio of gas pressure to total pressure $\beta=0.5$.
For accretion rate values above $\dot{M}_{crit}$ the only equilibrium
configuration available to the accretion flow is a standard thin disk. 
The region of the hot accretion flow for which non-local
cooling due to scattering of the cold photons emitted by the thin disk
is more efficient than local cooling processes is shaded.  When the
non-local cooling is included in the calculations, the critical accretion
rate decreases.  The new $\dot{M}_{crit}$ curve is plotted as a dashed 
heavy line. (b) Compares the temperature profile of the flow with (thin line)
and without (heavy line) extra cooling due to the inner thin disk. \\

\noindent 
Figure 4. The ratio of non-local cooling rate due to scattering of the 
radiation emitted in the outer thin disk, to the local cooling rate in the 
gas near the transition layer between the hot and cold flows, plotted 
as a function of the transition radius for two-temperature (a) and
one-temperature (b) models.  Different curves are labeled by the values of the
accretion rate in Eddington units.  Note that for $R_{\rm tr} \lsim 100 
R_{Schw}$, this ratio is significantly smaller in one-temperature models which 
have much stronger local cooling than their two-temperature counterparts.\\

\noindent
Figure 5. Critical mass accretion rate, $\dot{M}_{crit}$,
plotted in Eddington units (solid lines) for two-temperature models
with viscosity parameter $\alpha=0.1\ {\rm and}\ 0.3$.  The dashed lines
correspond to the minimum allowed values of the transition radius between the 
hot flow
and the outer thin disk.  If the transition occurs in the region to the left
of the dashed line, Compton cooling of the hot gas by the thin disk photons is
so strong that the entire flow settles to the thin disk configuration.

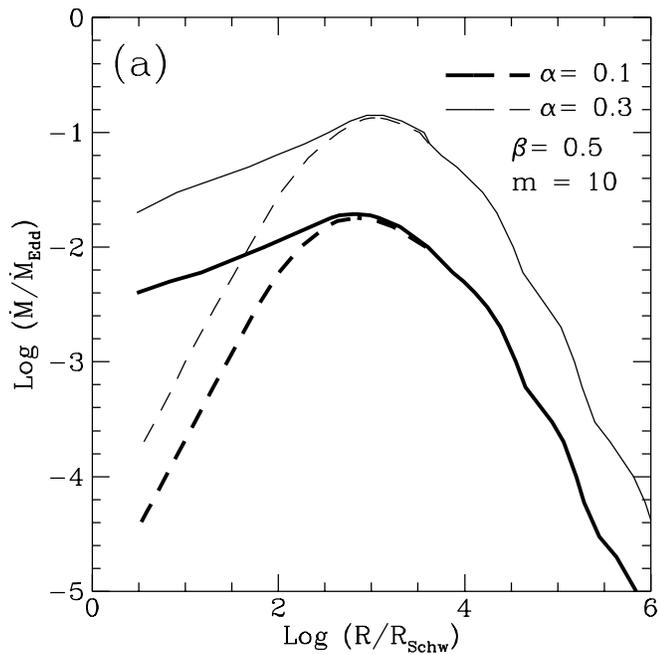
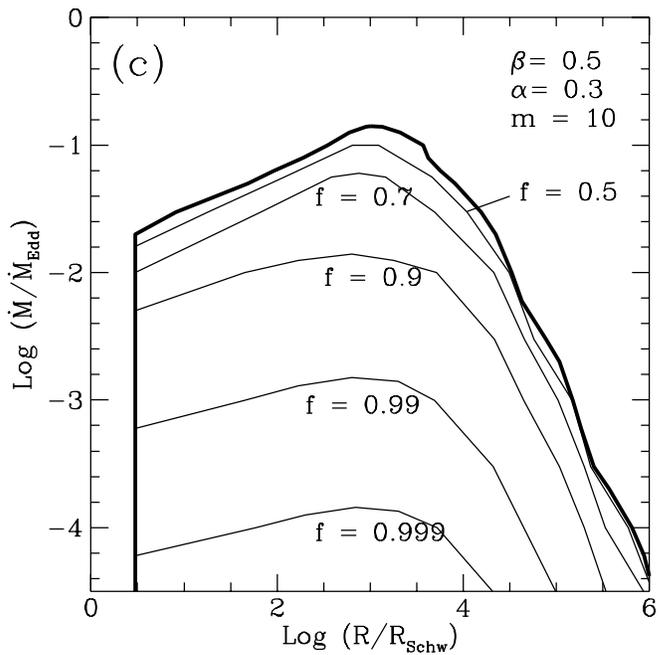
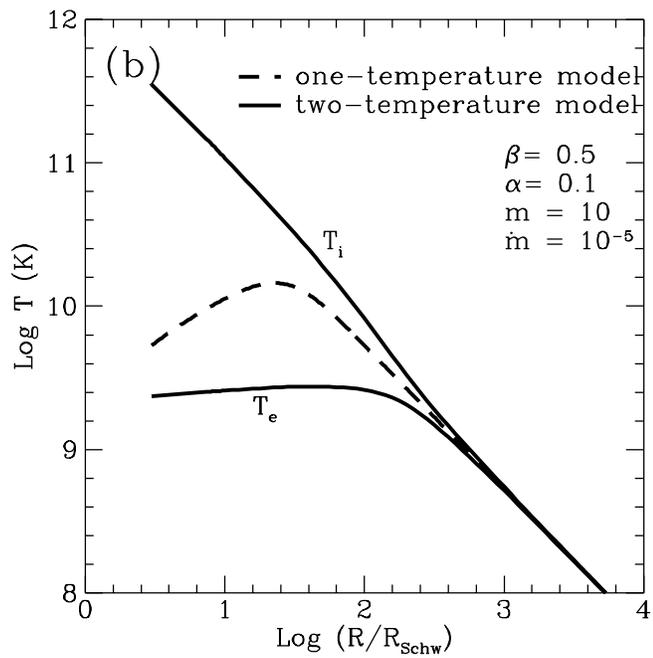
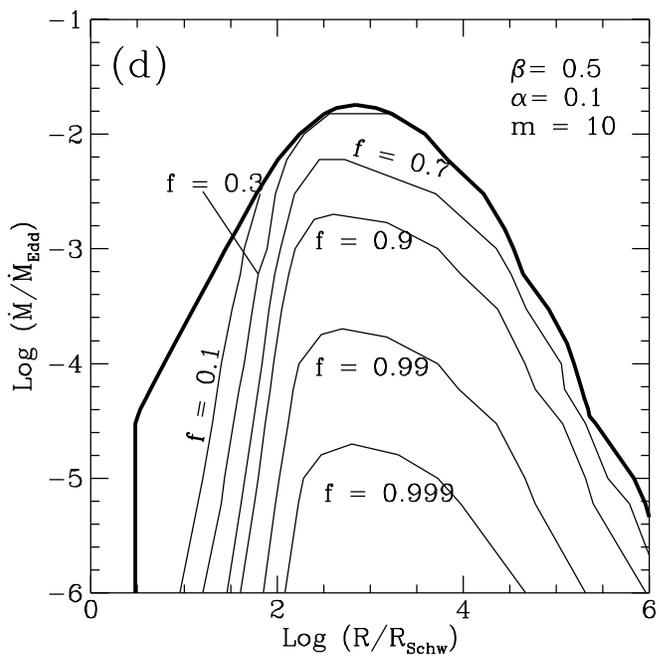
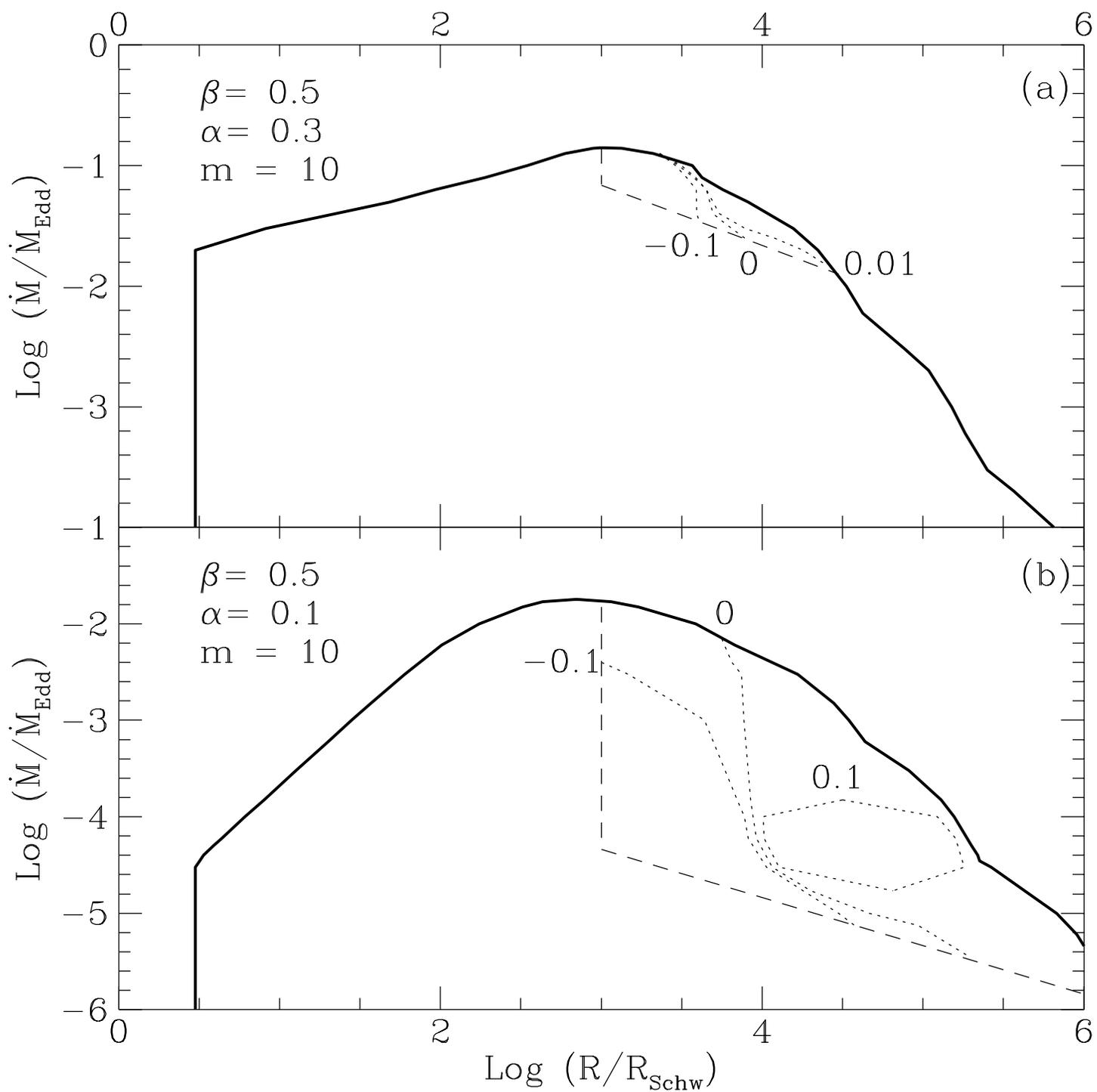
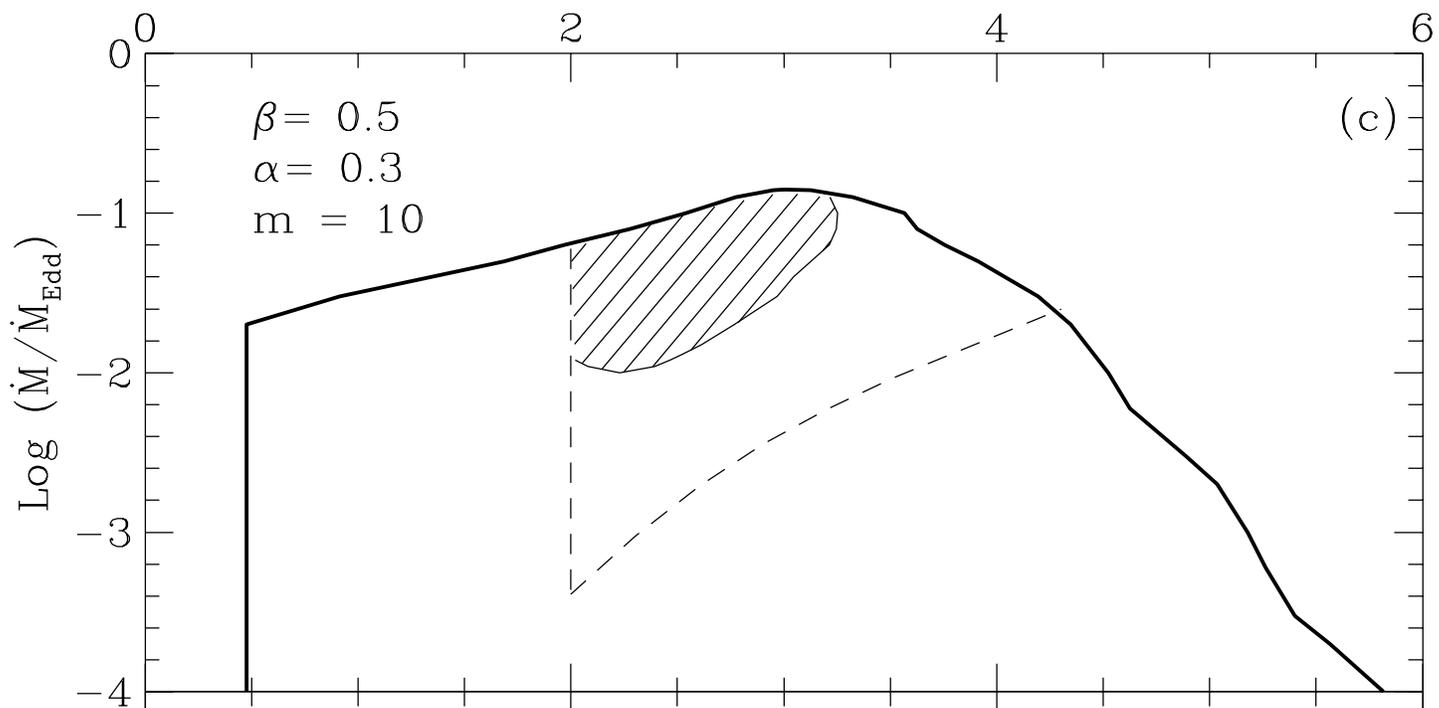
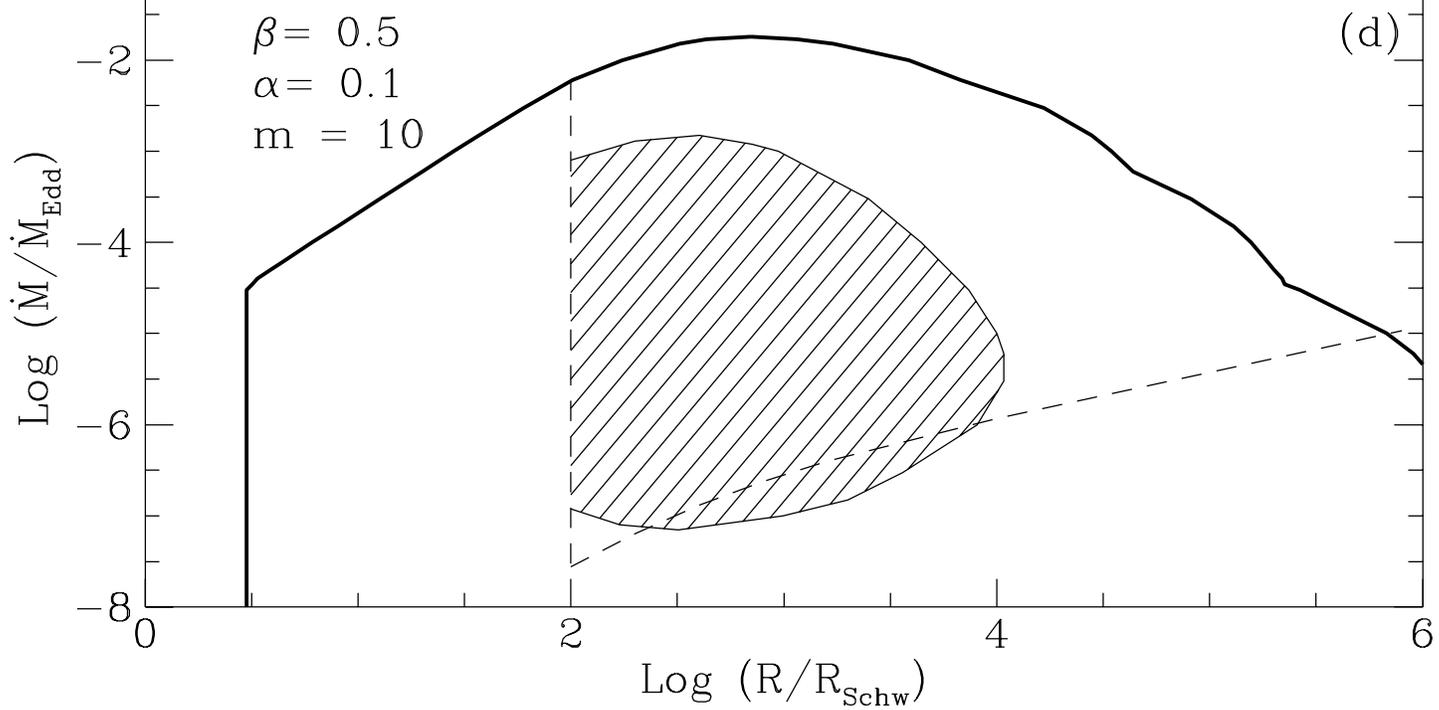
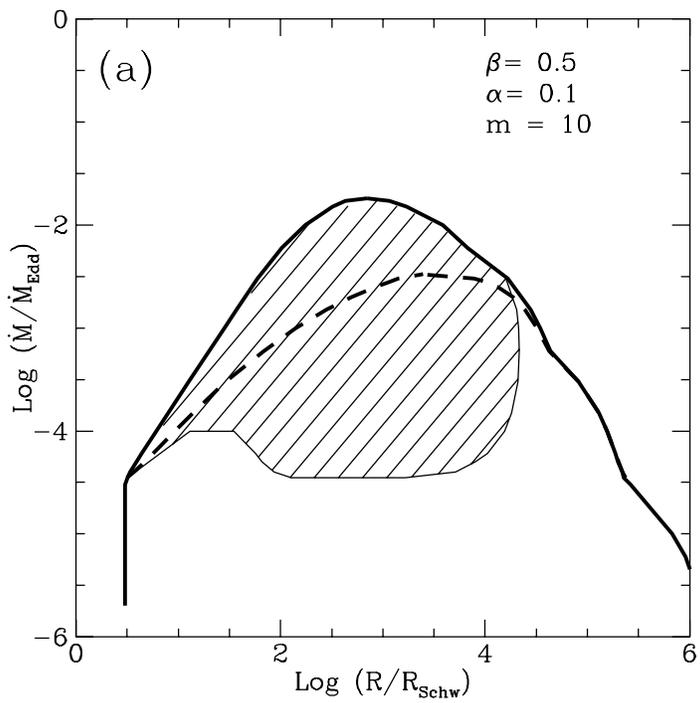
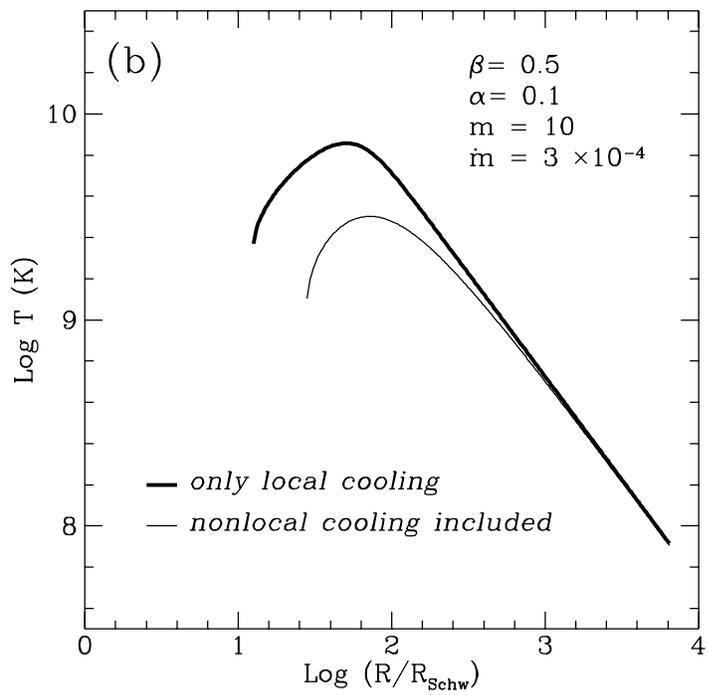
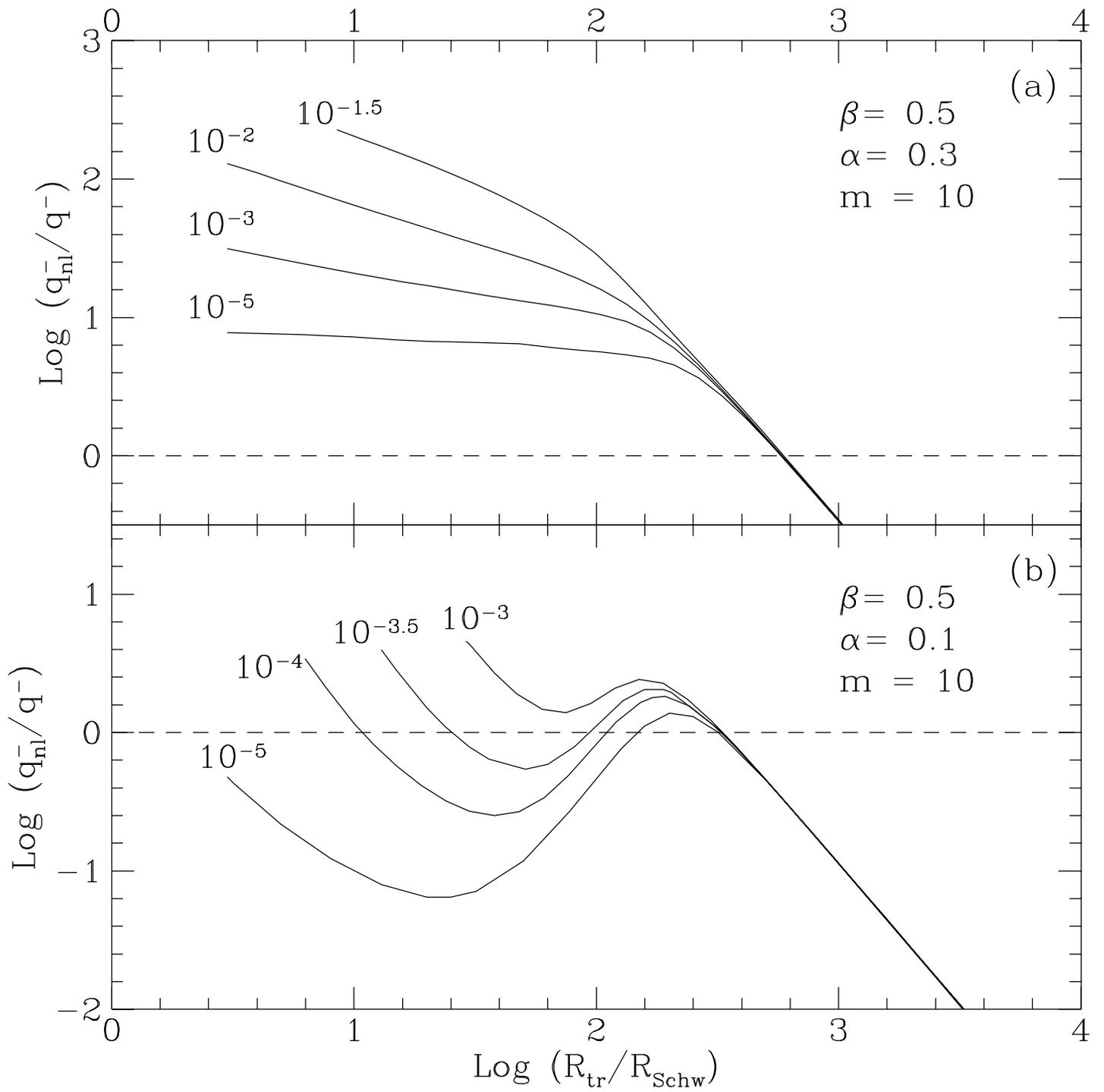
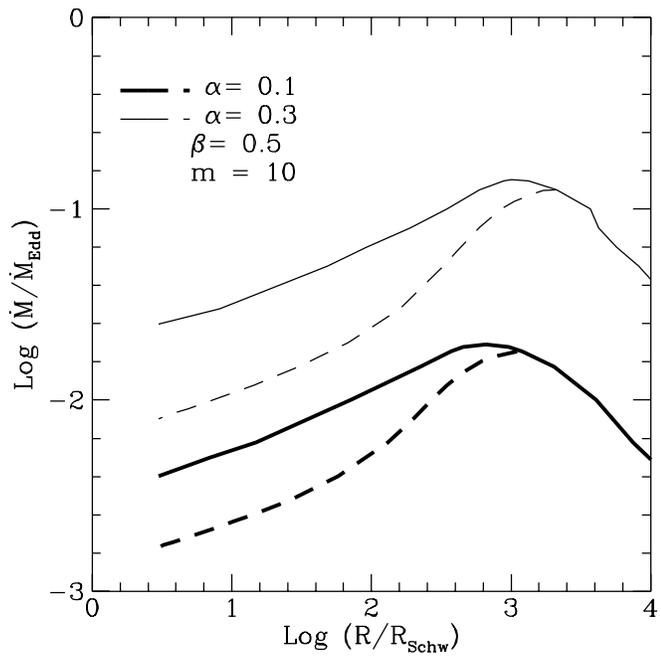
\end{document}